\documentclass[runningheads]{llncs}
\usepackage[T1]{fontenc}
\usepackage{graphicx}

\usepackage[symbol]{footmisc}
\usepackage{amsmath}
\usepackage{amssymb}
\usepackage{booktabs}
\usepackage{float}
\usepackage[misc,geometry]{ifsym}

\DeclareMathOperator*{\argmin}{arg\,min} 

\begin{document}
\title{Groupwise Deformable Registration of Diffusion Tensor Cardiovascular Magnetic Resonance: Disentangling Diffusion Contrast, Respiratory and Cardiac Motions}

\author{Fanwen Wang\inst{1,2}\Letter, Yihao Luo\inst{1}, Ke Wen\inst{2,3}, Jiahao Huang\inst{1,2}, Pedro F. Ferreira\inst{2,3}, Yaqing Luo\inst{2,3}, Yinzhe Wu\inst{1,2}, Camila Munoz\inst{2,3}, Dudley J. Pennell\inst{2,3}, Andrew D. Scott\inst{2,3}, Sonia Nielles-Vallespin\inst{2,3} and Guang Yang\inst{1,2,3,4}\Letter}
\authorrunning{Wang et al.}

\institute{
Bioengineering Department and Imperial-X, Imperial College London, UK 
\and 
Cardiovascular Magnetic Resonance Unit, Royal Brompton Hospital, Guy’s and St Thomas’ NHS Foundation Trust, UK \and
National Heart and Lung Institute, Imperial College London, UK \and
School of Biomedical Engineering and Imaging Sciences, King's College London, UK}
\maketitle         
\begin{abstract}
Diffusion tensor based cardiovascular magnetic resonance (DT-CMR) offers a non-invasive method to visualize the myocardial microstructure. With the assumption that the heart is stationary, frames are acquired with multiple repetitions for different diffusion encoding directions. However, motion from poor breath-holding and imprecise cardiac triggering complicates DT-CMR analysis, further challenged by its inherently low SNR, varied contrasts, and diffusion-induced textures. Our solution is a novel framework employing groupwise registration with an implicit template to isolate respiratory and cardiac motions, while a tensor-embedded branch preserves diffusion contrast textures. We've devised a loss refinement tailored for non-linear least squares fitting and low SNR conditions. Additionally, we introduce new physics-based and clinical metrics for performance evaluation. Access code and supplementary materials at: 
https://github.com/ayanglab/DTCMR-Reg

\end{abstract}
\section{Introduction}
Diffusion tensor based cardiovascular magnetic resonance (DT-CMR) uniquely visualizes in vivo myocardial microstructure by assuming a stationary heart and employing multiple repetitions across diffusion encoding directions to track water molecule motion~\cite{nielles2020cardiac}. Despite using breath-holding and cardiac triggering to mitigate physiological motion, prospective motion correction still necessitates image registration.

\begin{figure}[t]
\includegraphics[width=\textwidth]{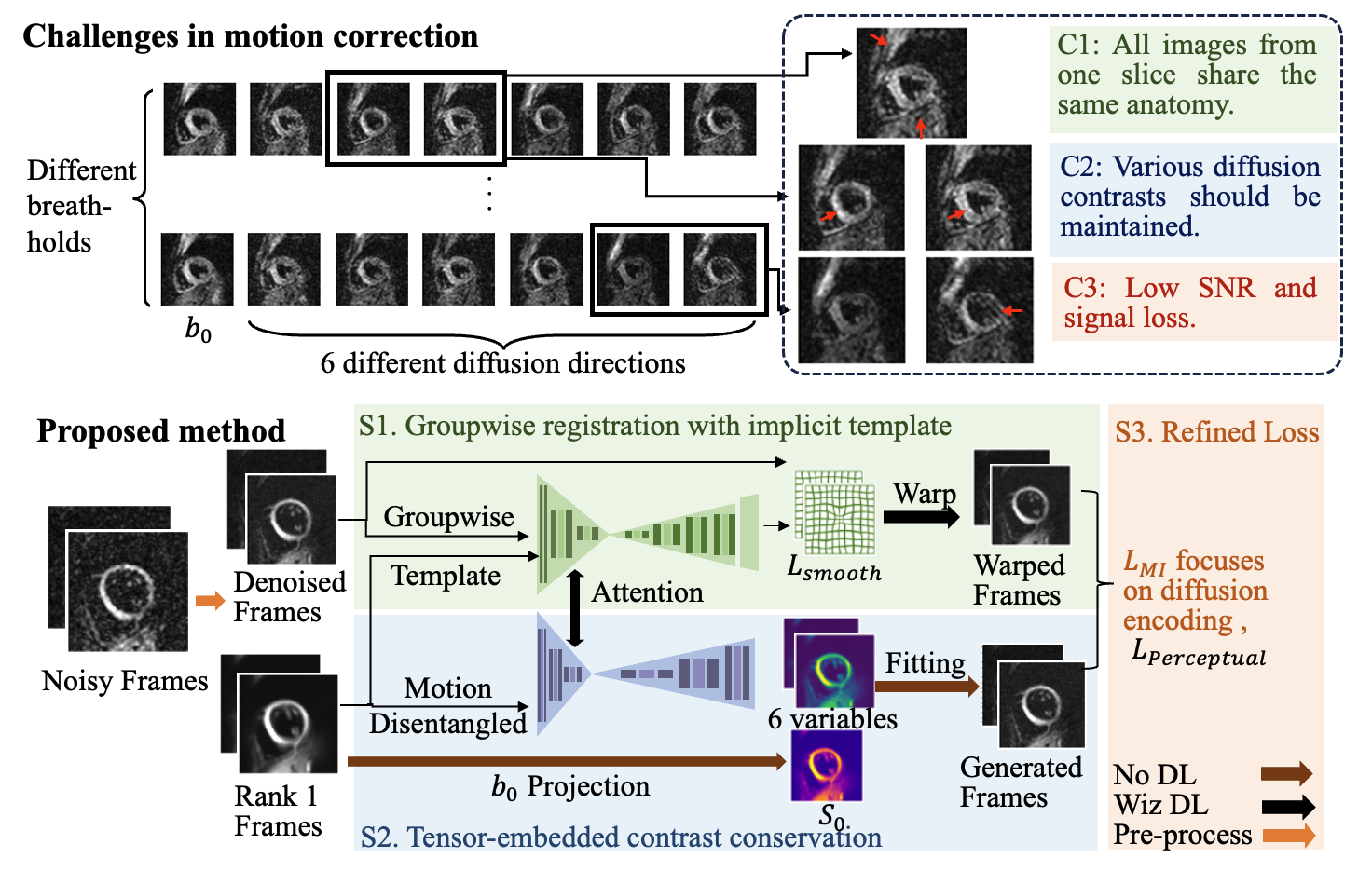}
\caption{The challenges in motion correction in DT-CMR and the proposed method with corresponding solutions. Given multiple scans of a single slice with varying contrasts (C1), our groupwise registration uses an implicit template to isolate motions (S1). To address diverse diffusion contrasts (C2), we integrated tensor data to maintain contrast and created pseudo frames for enhanced registration (S2). We further constrained our model with $S_0$ and $b_0$ information (S3). A unique differentiable mutual information loss is crafted for noise-reduced warped frames against generated ones (C3), fine-tuning our approach.} 
\label{DataAcquisitionAndMotionCorrection}
\end{figure}

Registration is a technique that aligns disparate images into a unified coordinate system, based on the intensity similarity or the theoretical likelihood. This criteria fails on DT-CMR since the textural information embedded should be retained for tensor fitting. Challenges amplify due to noise, varied diffusion contrasts, and surrounding tissues (Fig.~\ref{DataAcquisitionAndMotionCorrection}). Rigid registration is the most commonly applied method to correct for in-plane shifts but fail on the physiological motion. Nevertheless, cardiac and respiratory motions in the low SNR diffusion frames still call for deformable registration~\cite{ferreira2020automating}. 

In our study, we introduce an innovative registration technique for correcting non-rigid motion in DT-CMR acquisitions. To manage the challenge of acquiring a single slice through multiple repetitions with varied diffusion contrasts, we employ a group-wise registration strategy anchored by an implicit template~\cite{christodoulou2018magnetic}, which effectively separates respiratory motion and local deformations. To preserve diffusion contrast and encoding details, we incorporate tensor information matching the anatomy and generate pseudo-diffusion frames within a generation branch to facilitate registration. This approach, distinct from those in~\cite{hanania2023pcmc} and~\cite{zaffrani2022qdwi}, introduces additional tensor constraints to enhance frame fitting under low SNR conditions, leveraging differentiable mutual information as the primary loss function.

\begin{figure}[t]
\includegraphics[width=\textwidth]{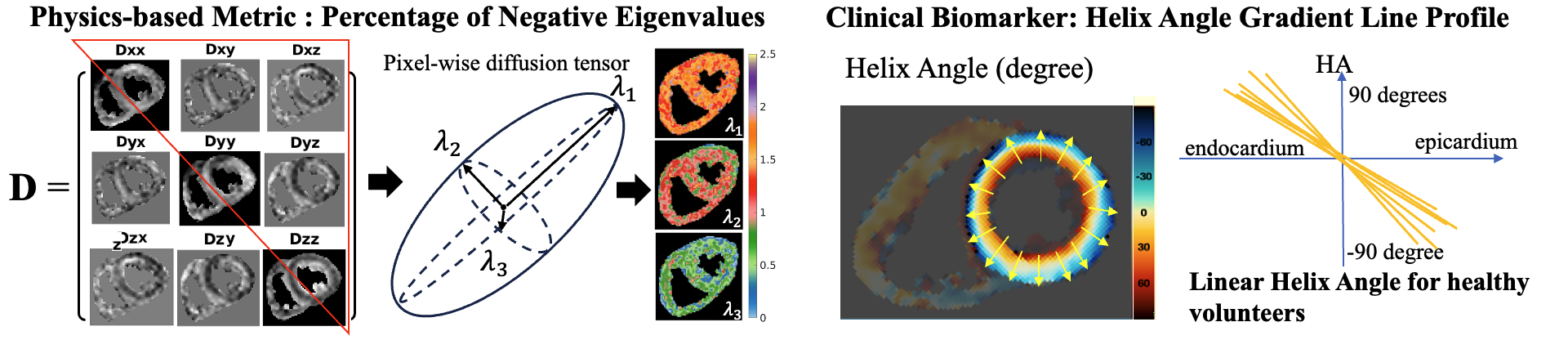}
\caption{Our evaluation introduces novel metrics grounded in physical properties and clinical markers. Each pixel's three eigenvalues, indicating water molecule motion, are expected to be inherently positive. The helix angle represents the cardiomyocytes' helical structure, which in healthy individuals, should display a linear gradient. To quantify this helical arrangement's consistency, we employ coefficient of determination (R$^2$) and root mean square error (RMSE) analyses of the helix angle gradient.}
\label{Evaluations}
\end{figure}

In Fig.~\ref{Evaluations}, we introduce two evaluations for DT-CMR: one quantifying physical integrity, and the other assessing clinical relevance. The first, the percentage of negative eigenvalues, measures noise-affected pixels, with valid diffusion tensors requiring three positive eigenvalues to reflect tissue water motion accurately. The second metric evaluates the consistency of cardiomyocyte orientation via the helix angle (HA). A linear gradient of this helical arrangement is indicative of a healthy heart, and non linearities may arise from residual motion and noise. To the best of our knowledge, this is the first physics-based metric and clinical biomarker designed to evaluate the performance of the registration.

Our proposed method shows superior performance when applied to DT-CMR assessed using two traditional and two deep-learning based methods. Novel evaluations on the tensor data also demonstrate the potential application of these methods. 

\section{Methods}
\subsection{Problem Formulation}
The fitting of diffusion images follows the pixel-wise Stejskal-Tanner equation~\cite{stejskal1965spin}: 
\begin{equation}
S_k = S_0 e^{-b_i \mathbf{g}_k^T \mathbf{D} \mathbf{g}_k},
\label{DTIEquation}
\end{equation}
where $S_k$ is the diffusion images we acquired, $S_0$ is the image intensity with no diffusion weighting and $b_i$ is the diffusion encoding strength. $\mathbf{D}$ is a symmetric 3$\times$3 covariance matrix with 6 independent variables (Fig.~\ref{DataAcquisitionAndMotionCorrection}). $\mathbf{g}_k$ is the diffusion encoding direction. 

We hypothesize that generating variables in the diffusion covariance matrix concurrently with motion correction will enhance the accuracy of the final tensor fitting. Therefore, we formulate the problem as follows:

\begin{equation}
\hat{\Phi}, \mathbf{D} = \argmin_{\Phi, \mathbf{D}} \sum_{i \in k} \mathcal{L}_{mi}(\phi_i \circ S_i - S_0 ^{-b_i \mathbf{g}_i^T \mathbf{D} \mathbf{g}_i}),
\end{equation}
where $\phi_i$ is the $i^{th}$ displacement field corresponding to the frame $S_i$. Specifically, $S_0$ is an extra constraint we put into the network. Six independent variables which stand for the upper triangular part of the diffusion covariance matrix in Fig.~\ref{Evaluations} can be derived from the network. 

\begin{table}[t]
\centering
\caption{Ablation studies over the whole test set. R$^2$ and RMSE are shown in (mean ± standard deviation). NE\% is shown in (median with 25\% and 75\% quantiles). SX stands for the modules in solution X.}
\begin{tabular}{lccc}
\hline
\textbf{Model} & \textbf{R$^2$ of HAG}($\uparrow$)& \textbf{RMSE of HAG}($\downarrow$)& \textbf{NE\%}($\downarrow$)\\
\hline
(S2.a) S0 constrain        & 0.894 ± 0.056 & 6.876 ± 3.605 & 1.283 (0.477, 3.259) \\
(S2.b) No attention        & 0.894 ± 0.057 & 6.978 ± 3.673 & 1.492 (0.695, 4.154) \\
(S3.a) No denoising        & 0.898 ± 0.054 & 6.556 ± 3.407 & 1.301 (0.690, 3.731) \\
(S3.b) MSE loss            & 0.890 ± 0.058 & 7.217 ± 3.699 & 1.761 (0.750, 4.585) \\
(S4.a) Pure seg            & 0.891 ± 0.057 & 7.213 ± 3.704 & 1.661 (0.766, 4.016) \\
(S4.b) +Seg                & 0.896 ± 0.055 & 6.758 ± 3.567 & 1.301 (0.818, 4.460) \\
Proposed & \bfseries{0.903 ± 0.053} & \bfseries{6.542 ± 3.423} & \bfseries{1.244 (0.519, 3.424)} \\
\hline
\end{tabular}
\label{table:Ablation Studies}
\end{table}

\subsection{Pipeline}
\subsubsection{Solution 1 Reference choice: Groupwise-based registration with implicit template}
In our group-wise registration approach using an implicit template, crucial for DT-CMR where one slice is sampled $N_{d} \times N_{rep}$ times ($N_{d}$: diffusion frames, $N_{rep}$: repetitions), we tackle the challenge of selecting an appropriate fixed reference for deformable registration. The implicit template is constructed in Tucker form~\cite{tucker1966some} as:
\begin{equation}
\mathbf{A} = \Phi_{xy} \cdot \Phi_{dynamic},
\end{equation}
with $\Phi_{xy}^{R}$ representing the image basis matrix and $\Phi_{dynamic}^{R}$ denoting the dynamic factor tensor ($N_{d} \times N_{rep}$), where R specifies the basis matrix rank. By setting R to 1, we generate a first-rank averaged projection tensor $\mathbf{A}^{1} = \Phi_{xy}^{1} \cdot \Phi_{dynamic}^{1}$, effectively separating respiratory and local deformations~\cite{nguyen2021free}. This template guides a b-spline registration network with diffeomorphic transformations to align frames and compute displacement fields~\cite{qiu2021learning}.

\subsubsection{Solution 2 Diffusion conservation: Tensor-embedded contrast conservation}
To conserve the textural information from diffusion encoding directions, we propose a tensor-embedded generation branch creating pseudo diffusion images with same anatomy but different contrast to guide the groupwise registration. Following Eq.~\ref{DTIEquation}, we use the averaged $1^{st}-$rank projection of tensor of $b_0$ frames as the $S_0$ . We avoid the blurring induced by misalignment of the original input images and take this as a further constraint. 
Using $\mathbf{A}^{1}$ as the input, six independent variables from the diffusion covariance matrix in Fig.~\ref{Evaluations} are generated from an encoder-decoder structure. The variables were then combined with $S{_0}$ in an exponential way following Eq.~\ref{DTIEquation} to generate pseudo-diffusion images. 
Besides, a cross-attention module is built between the generation and registration network in a forward manner for better connection. 

\subsubsection{Solution 3 Least-squares fitting: Refined loss on diffusion encoding information} 
To mitigate the impact of low SNR in diffusion images, which can adversely affect metrics and lead to significant deformation, we employ a strategy of extracting different components from $\Phi_{dynamic}$ across varied diffusion encodings and repetitions for initial denoising. This process is enhanced with an auto-correlation function to better preserve diffusion contrast~\cite{gurney2019principal}. Additionally, to address the challenges posed by substantial intensity fluctuations from motion or acquisition issues, and to align with the principles of least-squares tensor fitting, we opt for a differentiable mutual information loss $\mathcal{L}_{mi}$ as our primary loss metric, diverging from the traditional mean squared error approach used in previous studies~\cite{hanania2023pcmc,zaffrani2022qdwi}.The loss function consists of three parts:  
\begin{equation}
\mathcal{L}_{total} = \lambda_1 \cdot \mathcal{L}_{mi} + \lambda_2 \cdot \mathcal{L}_{smooth} + \lambda_3 \cdot \mathcal{L}_{perceptual},
\end{equation}
in which $\mathcal{L}_{mi}$ is defined on the image domain:
\begin{equation}
\mathcal{L}_{mi} = \sum_{i \in k}\frac{H(S_0 ^{-b_i \mathbf{g}_i^T \mathbf{D} \mathbf{g}_i}) + H(\phi_i \circ S_i)}{H(S_0 ^{-b_i \mathbf{g}_i^T \mathbf{D} \mathbf{g}_i},\phi_i \circ S_i)},
\end{equation}
where $H(S_0 ^{-b_i \mathbf{g}_i^T \mathbf{D} \mathbf{g}_i})$ and $H(\phi_i \circ S_i)$  represent the marginal entropies of the generated and the warped diffusion images and $H(S_0 ^{-b_i \mathbf{g}_i^T \mathbf{D} \mathbf{g}_i},\phi_i \circ S_i)$  represent the joint entropy. Using a Gaussian-based Parzen window~\cite{thevenaz2000optimization,qiu2021learning} instead of a rectangular window to calculate the intensity bins, we make the main loss differentiable and easy to back propagate. 

\paragraph{$\mathcal{L}_{smooth}$} is the $L2$ norm of the spatial gradient of the displacement field,
\begin{equation}
\mathcal{L}_{smooth} = \sum_{i \in k} \sum_{p \in \Omega}\|\nabla \phi_i(p)\|^2,
\end{equation}
where $\Omega$ is the 2-dimensional image area in our case. 

Furthermore, we incorporate a perceptual loss, $\mathcal{L}_{perceptual}$, to supervise the high-level feature similarity between pseudo-diffusion images and warped images, utilizing layers from the VGG16 model~\cite{simonyan2014very}.

\begin{figure}[t]
\includegraphics[width=\textwidth]{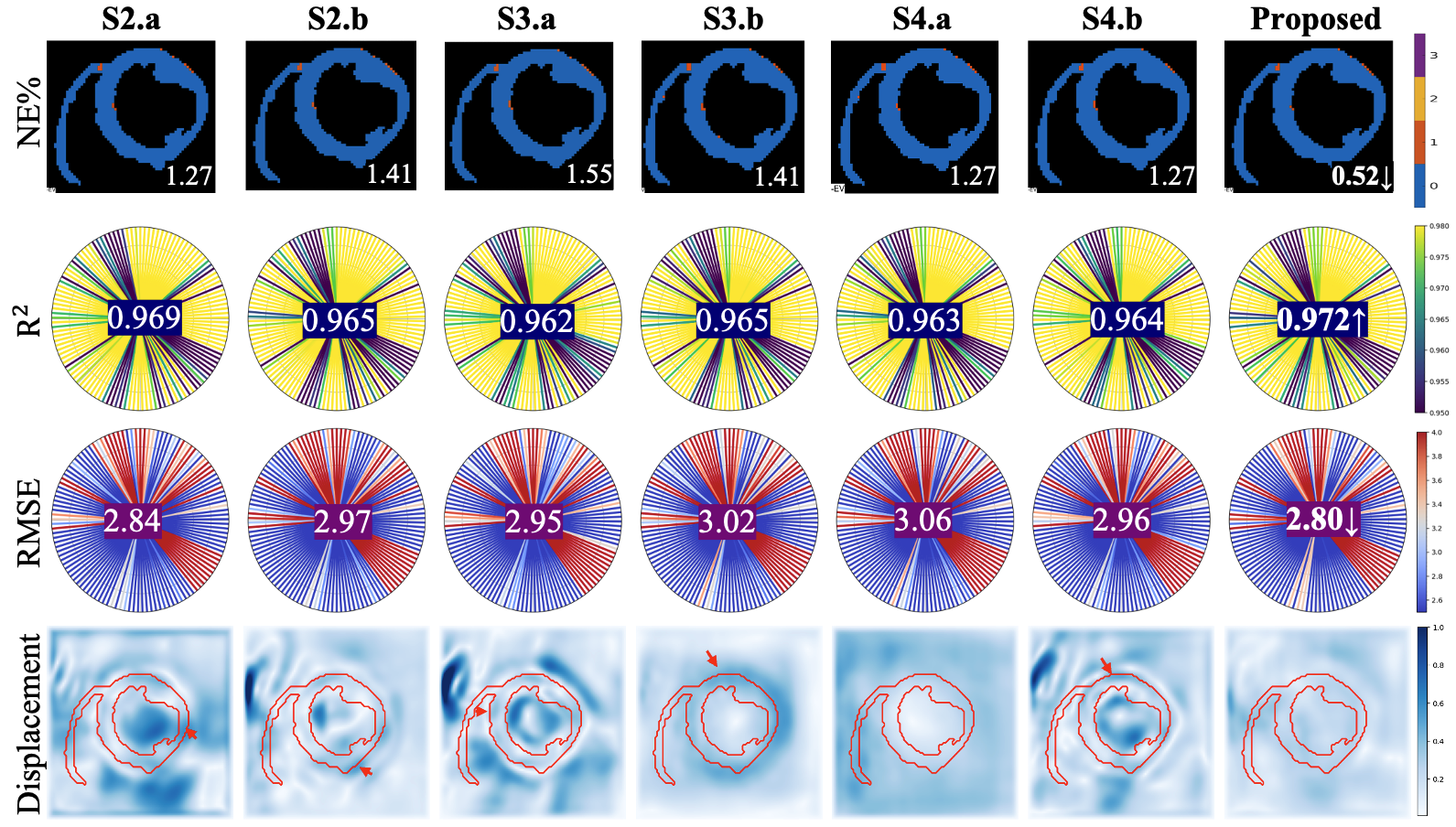}
\caption{The quantification of ablation studies with NE\% listed on the right bottom, R$^2$ and RMSE of HAG and displacement field along rows. SX stands for the solution X. S2.a stands for using MSE on $S_0$ as constraint; S2.b is without attention module; S3.a takes noisy images as input; S3.b uses MSE as the main loss; S4.a just uses DICE on the segmented label to supervise; S4.b combines the segmentation DICE and the mutual information loss together. The unfavourable drastic deformation is pointed out by red arrows.} 
\label{AblationVisual}
\end{figure}

\subsubsection{Solution 4 Contour guidance: Optional semi-supervised version}
A pre-trained segmentation network~\cite{ferreira2020automating} can be utilized to provide the mask of the myocardium as an auxiliary loss in a semi-supervised manner. We also tried to add a soft DICE between mask of the the intrinsic template of $S_0$ and the warped masks of different frames.

Notably, the displacement field $\phi_i$ inferred by the denoised input is applied on the original noisy frame in the following inference stage for a better and more robust fitting.

\subsection{Datasets}

Two datasets were collected from 20 healthy volunteers using Stimulated Echo Acquisition Mode (STEAM) based Echo Planar Imaging (EPI) at 3T and 1.5T scanners. \textbf{Dataset 1} comprises images at end-systole (ES) and end-diastole (ED) with a spatial resolution of 2.8 x 2.8 x 8.0 mm³, including 10 repetitions with $b_{0}$ and six high b-value encodings at b = 0 ($b_{0}$), 150 ($b_{150}$) and 600 ($b_{600}$) s/mm $^2$, resulting in 20 (subjects) $\times$ 2 (magnetic field) $\times$ 2(ES/ED) =80 cases. \textbf{Dataset 2} includes two short-axis slices (apical and basal) at ES, utilizing STEAM-EPI at 3T with twelve repetitions for $b_{0}$, $b_{150}$ and $b_{600}$. A number of 20 (subjects) $\times$ 2 (slices) = 40 cases are acquired after employing diaphragm positioning to reduce drastic through-plane motion artifacts. Hence, we got 120 cases in total. ECG-triggered breath-holding was applied during acquisitions. Data segmentation was performed using a pre-trained network~\cite{ferreira2020automating}, categorizing cases by segmentation quality. Eighty cases with superior segmentation were allocated into training (80\%) and validation (20\%) subsets, while the remaining 40, characterized by lower segmentation quality, were designated for testing. This setup aimed to challenge and evaluate model performance primarily on data with lower image quality.

\subsection{Experimental Details}
Traditional rigid~\cite{guizar2008efficient} and deformable registration method~\cite{klein2009elastix} were used for comparison. Two state-of-the-art (SOTA) deep learning based registration methods, namely MIDIR~\cite{qiu2021learning} and Transmorph~\cite{chen2022transmorph} were also compared. 

Ablation studies were performed to examine the various modules in our proposed method: (S2.a) MSE loss on $S_0$ instead of directly using the averaged projection of $b_{0}$, (S2.b) no cross-attention module, (S3.a) noisy frames as input, (S3.b) MSE as loss, segmentation label incorporation with (S4.a) purely soft DICE on segmentation labels, (S4.b) combination of differentiable MI on the frames and soft DICE on the segmented labels, and the proposed. Details about the implementation and hyper parameter settings can be found in supplementary material.

\subsection{Evaluation Methods}
We compacted all the frames along a third dimension to examine the alignment of image stack. Besides, two extra quantification were applied as in Fig.~\ref{Evaluations}: 
\subsubsection{Helix Angle Gradient (HAG) Line Profile} In healthy volunteers, the helix angle (HA) reveals the alignment of cardiomyocytes, which ideally should be linear from the endocardium to the epicardium, as shown in Fig.~\ref{Evaluations}. The mean HAG was calculated as degree of the alignment over the percentage of myocardial wall thickness. Radial line profiles with linear regression showing negative slopes and a fitting R$^2$ greater than 0.3 were included for analysis~\cite{gorodezky2018diffusion}. The R$^2$ values and RMSE of these fittings were evaluated. 
\subsubsection{Percentage of Negative Eigenvalues (NE\%)} As illustrated in Fig.~\ref{Evaluations}, a rank-2 diffusion tensor and its eigenvalues were calculated for each pixel of the diffusion dataset~\cite{ferreira2014vivo} These three eigenvalues indicate the magnitude of water molecule movement, which should ideally all be positive. However, noise, residual motion or misregistration artifacts may increase the NE\%. We calculated the NE\% in the left ventricular myocardium. 

\begin{figure}[t]
\includegraphics[width=\textwidth]{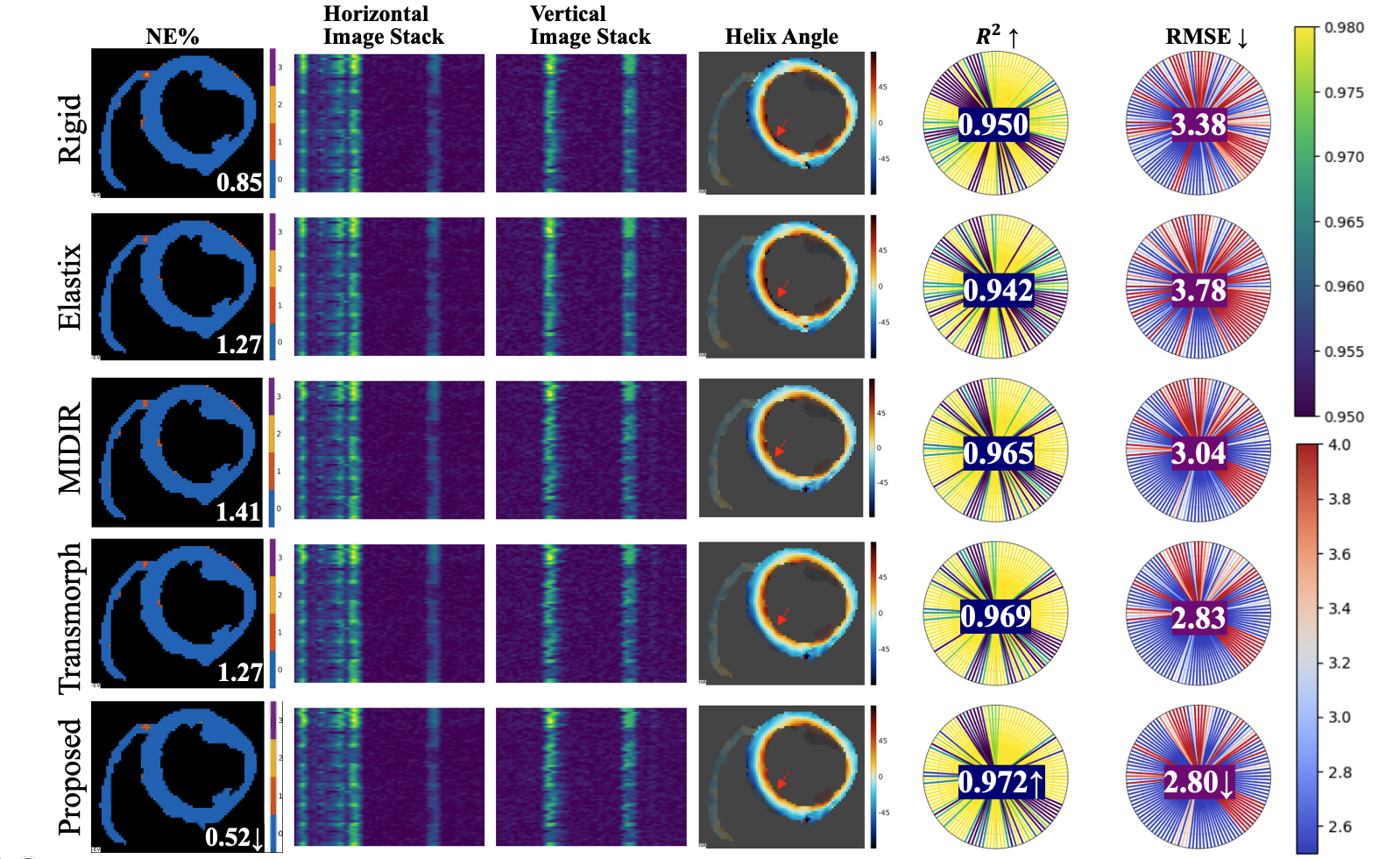}
\caption{The result of tensors of different comparison studies on one representative subject. NE\%, image stack, R$^2$ and RMSE of HAG are shown. Differences along the edge of HA are highlighted by red arrows.} 
\label{CompaVisual}
\end{figure}

\begin{table}[t]
\centering
\caption{Comparison studies over the whole test dataset. R$^2$ and RMSE are shown in (mean ± standard deviation). NE\% is shown in (median with 25\% and 75\% quantiles)}
\begin{tabular}{lccc}
\hline
\textbf{Model} & \textbf{R$^2$ of HAG}($\uparrow$)& \textbf{RMSE of HAG}($\downarrow$)& \textbf{NE\%}($\downarrow$)\\
\hline
Rigid & 0.873 ± 0.061 & 7.583 ± 3.739 & 1.251 (0.546, 4.078)  \\
Deformable & 0.881 ± 0.053 & 7.502 ± 3.544 & 1.459 (0.439, 4.136) \\
MIDIR & 0.889 ± 0.059 & 7.266 ± 3.740 & 1.727(0.790, 4.835) \\
Transmorph & 0.892 ± 0.057 & 7.144 ± 3.649 & 1.703 (0.508, 4.259)\\
Proposed & \bfseries{0.903 ± 0.053} & \bfseries{6.542 ± 3.423} &  \bfseries{1.244 (0.519, 3.424)} \\
\hline
\end{tabular}
\label{table:Comparison Studies}
\end{table}

\section{Results}
\subsection{Ablation and Comparison Study} 

In our study, we quantitatively compared various modules' performance on a representative subject, as shown in Fig.\ref{AblationVisual}. The proposed method, relying exclusively on the loss of generated frames, surpassed others, demonstrated alongside an example displacement field. Issues such as drastic myocardium deformation, highlighted by red arrows, were observed with strategies like applying MSE to $S_0$, using noisy frames as inputs, omitting the attention module, employing MSE loss, and integrating segmentation labels. These findings are further substantiated in Table~\ref{table:Ablation Studies}.

Additionally, our approach was benchmarked against conventional and deep-learning based techniques, detailed in Fig.~\ref{CompaVisual} and Table~\ref{table:Comparison Studies}. It consistently outperformed competing methods, showcasing the lowest percentage of negative eigenvalues, highest R$^2$, and minimal RMSE for helix angle gradient linear fitting, indicating superior performance across metrics for a representative subject.

\section{Conclusion and Discussion}
In this study, we introduce a groupwise registration method for DT-CMR to address challenges posed by multiple repetitions, diverse contrasts, and noise. An implicit template segregates respiratory and cardiac motions for group-wise frame registration. Our tensor-embedded network generates pseudo frames to differentiate diffusion contrasts, while a novel differentiable mutual information loss targets diffusion encoding to correct residual motions.

We validated our method with two new metrics, physical and clinical, demonstrating its superiority over traditional and deep learning-based registration methods. Our approach significantly reduces negative eigenvalues and improves the helix angle gradient in healthy volunteers. However, the proposed clinical biomarker of the helix angle gradient line profile is currently applicable only to healthy volunteers due to the linear features of cardiomyocytes in these individuals. To address this limitation, we plan to include additional metrics and expand our study to patients with various diseases in future research. This will enhance the applicability and robustness of our method.

\begin{credits}

\subsubsection{\ackname}This work was supported in part by ERC IMI (101005122), the H2020 (952172), the MRC (MC/PC/21013), the Royal Society (IEC\textbackslash NSFC\textbackslash 211235), the NVIDIA Academic Hardware Grant Program, the SABER project supported by Boehringer Ingelheim Ltd, Wellcome Leap Dynamic Resilience, the UKRI Future Leaders Fellowship (MR/V023799/1) and the British Heart Foundation grant (RG/19/1/34160).

\subsubsection{\discintname} We have no competing interests in the paper as required by the publisher.

\end{credits}

\clearpage
\bibliographystyle{splncs04}
\bibliography{ref}
%
\end{document}